\documentclass[aps,showpacs,prb,twocolumn]{revtex4}
\usepackage{graphicx}

\begin{document}

\title{Pressure-tuned First-order Phase Transition and Accompanying Resistivity Anomaly in CeZn$_{1-\delta}$Sb$_{2}$}
\author{Tuson Park$^{1}$, V. A. Sidorov$^{1,2}$, Hanoh Lee$^{3}$, Z. Fisk$^{3}$, and J. D. Thompson$^{1}$}
\affiliation{$^{1}$Los Alamos National Laboratory, Los Alamos, New Mexico 87545, USA \\
$^{2}$Vereshchagin Institute of High Pressure Physics, 142190 Troitsk, Russia\\
$^{3}$Department of Physics, University of California at Davis, CA 95616, USA}
\date{\today}

\begin{abstract}
The Kondo lattice system CeZn$_{0.66}$Sb$_{2}$ is studied by electrical resistivity and ac magnetic susceptibility measurements at several pressures. At $P=0$~kbar, ferromagnetic and antiferromagnetic transitions appear at 3.6 and 0.8~K, respectively. The electrical resistivity at $T_N$ dramatically changes from the Fisher-Langer type (ferromagnetic like) to the Suezaki-Mori type near $17$~kbar, i.e., from a positive divergence to a negative divergence in the temperature derivative of the resistivity. The pressure-induced SM type anomaly, which shows thermal hysteresis, is easily suppressed by small magnetic field (1.9~kOe for 19.8~kbar), indicating a weakly first-order nature of the transition. By subtracting a low-pressure data set, we directly compare the resistivity anomaly with the SM theory without any assumption on backgrounds, where the negative divergence in $d\rho/dT$ is ascribed to enhanced critical fluctuations in the presence of superzone gaps.
\end{abstract}
\maketitle

Rare-earth and actinide compounds with localized $f$-electrons at high temperatures show non-magnetic or even a superconducting ground state, which cannot be explained by Hill's observation that magnetic ordering depends on the $f$-atom spacing: magnetism at large $f-f$ spacing and paramagnetism or superconductivity at small separation.\cite{hill70} The importance of $s-f$ exchange coupling $J$ between localized $f$-electrons and conduction band electrons for determining the ground state, which was neglected by Hill, is typified in heavy-electron, Kondo-lattice, or mixed-valence systems, where a paramagnetic or a superconducting ground state is observed even though they have large $f-f$ spacing.\cite{fisk88,thompson92} Depending on the hybridization strength $|JN_{F}|$, the ground state is determined through competition between Kondo and Ruderman-Kittel-Kasuya-Yosida (RKKY) interactions, where $N_{F}$ is the conduction band density-of-states at the Fermi energy $E_{F}$. For $|JN_{F}|<< 1$, a magnetic state is stabilized because the RKKY interaction that provides coupling between local moments depends geometrically on $|J|$ - $T_{RKKY} \propto J^{2}N_{F}$, whereas a Kondo singlet is preferred for large $JN_{F}$ because the Kondo interaction that screens local moments depends exponentially on $|J|$ - $T_{K} \propto \text{exp}(-1/|J|N_{F})$. \cite{doniach77}
  
Another important consequence of the exchange interaction between the localized and conduction electrons is a change in the electronic structure. When an antiferromagnetic structure (AF) with a period incommensurate with the ionic lattice appears, the magnetic superlattice may distort the Fermi surface dramatically,\cite{mackintosh62} forming magnetic superzone gaps below $T_N$ when the ordering wavevector $K_A$ connects portions of the Fermi surface.\cite{salamon72} Suezaki and Mori (SM) showed that, when combined with enhanced spin scattering in a $K=K_A$ mode in antiferromagnets, this sharp band gap gives rise to a sharp increase in the resistivity or a negative divergence in the temperature derivative near $T_N$.\cite{suezaki69} Early measurements on rare-earth metals and order-disorder systems revealed a similar resistivity anomaly, but quantitative analysis has been limited due to the smearing of the gaps by thermal phonons and temperature dependent backgrounds.\cite{craven73,thomas73} Here we report a pressure-tuned first-order phase transition and an accompanying negative divergence in the temperature derivative of the resistivity of CeZn$_{0.66}$Sb$_{2}$. The first-order, SM type transition at $T_{N}$ only appears at intermediate pressures ($17.3\leq P < 25.5$~kbar), while the transition shows a Fisher-Langer type anomaly, i.e., a positive divergence in $d\rho / dT$ in the low pressure limit and a slight slope change at $P\geq25.5$~kbar. The first-order anomaly is suppressed by a magnetic field as small as 1.9~kOe at 19.8~kbar, indicating the transition is very weakly first order. By subtracting a low-pressure data set, we directly compare the resistivity anomaly with the SM theory without any assumption on backgrounds, where the negative divergence in $d\rho/dT$ is ascribed to enhanced critical fluctuations in the presence of superzone gaps.

CeZn$_{0.66}$Sb$_{2}$ was grown with Sb self flux in an evacuated and sealed quartz ampule and crystallizes in the tetragonal ZrCuSi$_2$ structure with space group P4/nmm. \cite{hanoh05} The electrical resistivity of CeZn$_{0.66}$Sb$_{2}$ at 0.3~K is $4.9~\mu \Omega \cdot$cm and the resistivity ratio is $\rho (=300\text{K})/\rho(=0.3\text(K))\approx 14$. Hydrostatic pressure up to 25~kbar was achieved by using a hybrid Be-Cu/NiCrAl clamp-type pressure cell with silicon fluid as a transmitting medium. At higher pressures, a profiled toroidal anvil clamped device was used with anvils supplied with a boron-epoxy gasket and a teflon capsule filled with glycerol/water mixture with volume ratio 3:2. Superconducting transition temperatures of Tin and Lead were used to determine pressure for the clamp-type and the toroidal anvil cell, respectively. The width of the superconducting transition is independent of pressure and is less than 10~mK up to 55~kbar, indicating that measurements were performed in hydrostatic conditions. Electrical resistivity $\rho$ was measured by a standard four-point method with a LR-700 ac resistance bridge (Linear Research) for current flowing perpendicular to the c-axis of CeZn$_{0.66}$Sb$_{2}$ ($I~\bot$~c-axis). AC magnetic susceptibility $\chi _{\text{ac}}$ in the plane was measured at $f=157$~Hz by a conventional method using primary and secondary pick-up coils mounted inside the pressure cell.

\begin{figure}[tbp]
\centering  \includegraphics[width=8cm,clip]{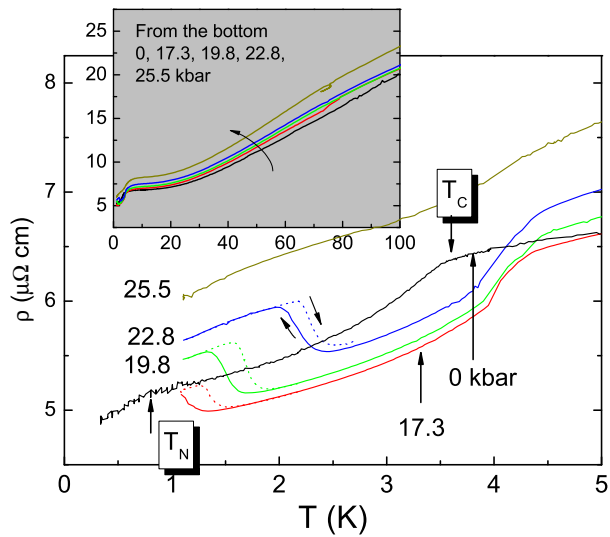}
\caption{(color online). Resistivity versus temperature at 0, 17.3, 19.8, 22.8, and 25.5~kbar for $I~\bot$~c-axis. Dotted lines are data taken with increasing temperature and solid lines, with decreasing temperature. Inset: $\rho$ vs. $T$ in more extensive temperature range.}
\label{figure1}
\end{figure}
Figure~1 shows the resistivity of CeZn$_{0.66}$Sb$_{2}$ as a function of temperature at several pressures. At ambient pressure, a sharp decrease occurs at 3.6~K, corresponding to a ferromagnetic phase transition observed in specific heat measurements. \cite{hanoh05} With increasing pressure, the ferromagnetic transition temperature ($T_{c}$) increases at a rate $dT_{c}/dP\approx 0.05$~K/kbar and the resistivity above $T_{c}$ slightly increases (see inset to Fig.~1), which could be explained by the enhanced hybridization between $f-$ and itinerant electrons with pressure. A slight decrease in $\rho$ is also observed at 0.85~K and ambient pressure, which corresponds to an antiferromagnetic transition observed in Ref.~[10]. The entropy recovered at $T_{N}$ is about 20~\% of $Rln2$,\cite{hanoh05} suggesting that the AF state at low temperature is a bulk property, not due to an impurity phase. The antiferromagnetic transition temperature $T_{N}$ slowly increases with pressure, $dT_{N}/dP \approx 0.02$~K/kbar. At 17.3~kbar, a peak-like feature with thermal hysteresis appears: the resistivity sharply increases with decreasing temperature, and then decreases. As shown in Fig.~1, with further increasing pressure, the anomaly becomes more pronounced and the transition temperature increases at a much faster rate, $dT_{N}/dP \approx 0.14$~K/kbar. Above 25.5~kbar, however, the peak disappears and only a slight slope change occurs at $T_N$. A similar pressure-induced resistivity peak was reported in CeRhGe. \cite{asai03}

Figure~2 shows representative resistivity data at 19.3~kbar as a function of temperature for several fields. Even though a different piece was used for this measurement, the resistivity anomaly is still reproducible, indicating it is intrinsic. The resistivity peak and thermal hysteresis are steeply suppressed with magnetic field. At as small a magnetic field as 1.9~kOe within the ab-plane, it is totally depressed, suggesting that the AF phase transition is weakly first order. The inset to Fig.~2 gives the temperature derivative of the resistivity. When there is a negative divergence in $d\rho / dT$, we assigned the divergent point as the transition temperature $(T_{N})$, while the maximum point was assigned to $T_{N}$ for the field without negative divergence. $T_{N}$ decreases with magnetic field, $dT_{N}/dH \approx -0.13 \pm 0.04$~K/kOe.
\begin{figure}[tbp]
\centering  \includegraphics[width=8cm,clip]{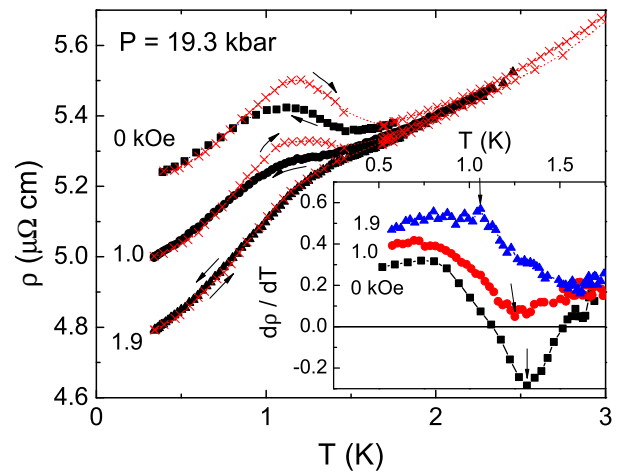}
\caption{(color online). Resistivity versus temperature for several magnetic fields at $P=19.3$~kbar for $H_{dc}$, $I~\bot$ c-axis. Inset: temperature derivative of the resistivity for cooling curve. Arrows indicate the transition temperature $T_{N}$ (see text).}
\label{figure2}
\end{figure}

Figure~3(a) shows the evolution of the temperature derivative of the resistivity for several pressures. At the ferromagnetic transition temperature, $d\rho /dT$ diverges for $T\rightarrow T_{c}^{+}$. We fit the positive divergence to the following form that is commonly used for critical fluctuations for $T>T_{c}$:
\begin{equation}
d\rho/dT = (A/\epsilon)(1+|t|^{-\epsilon})+B,
\end{equation}
where $t=(T-T_{c})/T_{c}$. When $\epsilon$ approaches zero, the above form suggests a logarithmic singularity at $T_{c}$. The inset to Fig.~3(a) magnifies $d\rho /dT$ for 19.8~kbar near $T_{c}$ and the solid line is the least-square fit to Eq.~(1). The best result was obtained with $T_{c}=4.419$~K and $\epsilon=0.04$, where the critical exponent is similar to other ferromagnets.\cite{zumsteg70} However, as pointed out by Kadanoff et al.,\cite{kadanoff67} the determination of the critical exponent depends on the range of the fit and availability of a number of data points near $T_{c}$. Nevertheless, a similar sharp peak, like that shown in the inset, is observed in the specific heat, \cite{hanoh05} which could be explained by the Fisher-Langer prediction that the magnetic contributions to $d\rho / dT$ and the specific heat of a ferromagnet should be proportional because short-range spin-correlations dominate in the temperature dependence of both quantities.\cite{fisher68} Below $T_{c}$, the AF magnetic transition makes it difficult to analyze the critical behavior. Above 25.5~kbar where the resistivity anomaly disappears, the peak in $d\rho /dT$ becomes broadened, making a quantitative fit impossible.
\begin{figure}[tbp]
\centering  \includegraphics[width=8cm,clip]{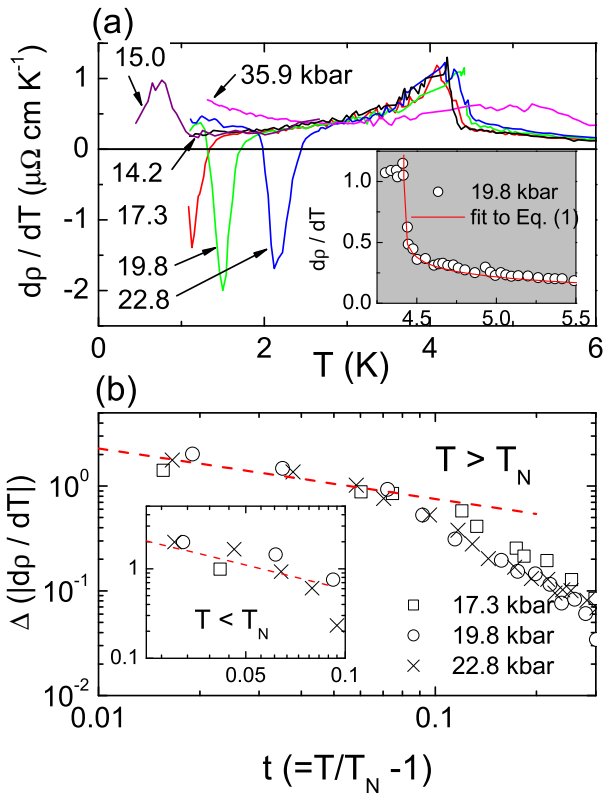}
\caption{(color online). (a) Temperature derivative of resistivity for cooling curve at 14.2, 15.0, 17.3, 19.8, 22.8, and 35.9~kbar. Inset: Positive divergence in $d\rho /dT$ of 19.8~kbar at $T_c$ is fitted to Eq.~(1). (b) $\Delta |d\rho /dT|$ versus $t$ at 17.3, 19.8, and 22.8~kbar. The negative divergence for both $T>T_N$ and $T<T_N$ (inset) is compared with the Suezaki-Mori theory (see text).}
\label{figure3}
\end{figure}

The negative divergence in $d\rho /dT$ at the N$\acute{\text{e}}$el temperature for intermediate pressures (Fig.~3a) can be understood in terms of combined effects of AF critical fluctuations and superzone gaps below $T_{N}$. In electrical resistivity $\rho = m/e^{2}\tau n_{eff}$, the effective number of carriers $n_{eff}$ depends on superzone gaps arising from the additional magnetic lattice periodicity, while the scattering rate $1/\tau$ is related to critical scattering of conduction electrons by localized spins. In ferromagnets, spin fluctuations with wave vectors close to $K_{A}=0$ contribute to small angle scattering of conduction electrons, leading to a weak anomaly, i.e., Fisher-Langer type. In antiferromagnets including those with a helical structure, on the other hand, critical fluctuations around the ordering wave vector $K_{A}=Q$ contribute to large angle scattering, leading to a large anomaly. Suezaki and Mori (SM) took into account this critical scattering and predicted the following form:\cite{suezaki69}
\begin{equation}
\begin{array}{ll}
d\rho /dT =-Bt^{-1/3}, \hspace{1.5cm} (T>T_{N}),\\
\hspace{1.07cm}=B^{'}t^{-1/3}-B_{g}t^{-2/3}, \hspace{0.2cm} (T<T_{N}).
\end{array}
\end{equation} 
For $T<T_{N}$, the first term is due to critical fluctuations and the second term is from long-range order, while only critical fluctuations contribute to the resistivity for $T>T_N$. Direct comparison of the critical phenomena between experiments and theory has been limited due to other contributions to the temperature dependent resistivity, such as lattice vibrations. In order to account for other contributions, we subtracted a low-pressure data set (14.2~kbar) because $T_c$ is close to that at intermediate pressures even though $T_N$ is below 1~K. Figure~3(b) shows  representative $\Delta |d\rho /dT|$ for $T>T_N$ as a function of $t (=T/T_{N}-1)$ for intermediate pressures, where  $\Delta |d\rho /dT|=|d\rho / dT (P)-d\rho/dT (14.2~\text{kbar})|$ and $T_N$ is assigned as the negative peak in $d\rho /dT$. $\Delta |d\rho /dT|$ for intermediate pressures shows scaling behavior, indicating the validity of the background subtraction. For direct comparison with the SM theory, a power-law form, $\Delta d\rho /dT = -Bt^{-\alpha}$, was used for $T>T_N$ and the best result was obtained with $B=0.25$ and $\alpha = 0.48\pm 0.08$ by least-squares technique (dashed line in Fig.~3b). The obtained exponent is compatible with the predicted value 1/3 from the SM theory. For $T<T_{N}$, all data sets, similar to those for $T>T_N$, collapse on top of each other and Eq.~(2) gives a good description of the data with $B^{'}=0.37$ and $B_g=0.32$ (see inset to Fig.~3b). We note that our analysis suffers from the limited temperature range of fitting because the low AF transition temperature makes it difficult to access a reasonable reduced temperature range near $T_N$: $t=0.01$ for $\Delta T =T-T_{N}\approx 10$~mK. As in Cr where a larger exponent was obtained,\cite{akiba72} the weakly first order nature of the transition can also complicate the analysis. The relatively good agreement between the experimental data and the SM theory in CeZn$_{0.66}$Sb$_2$, however, suggests that the resistivity anomaly mainly comes from critical fluctuations and magnetic gaps , which is consistent with the conclusion from the magnetic field dependence of the anomaly that the resistive transition is weakly first order. 

\begin{figure}[tbp]
\centering  \includegraphics[width=8cm,clip]{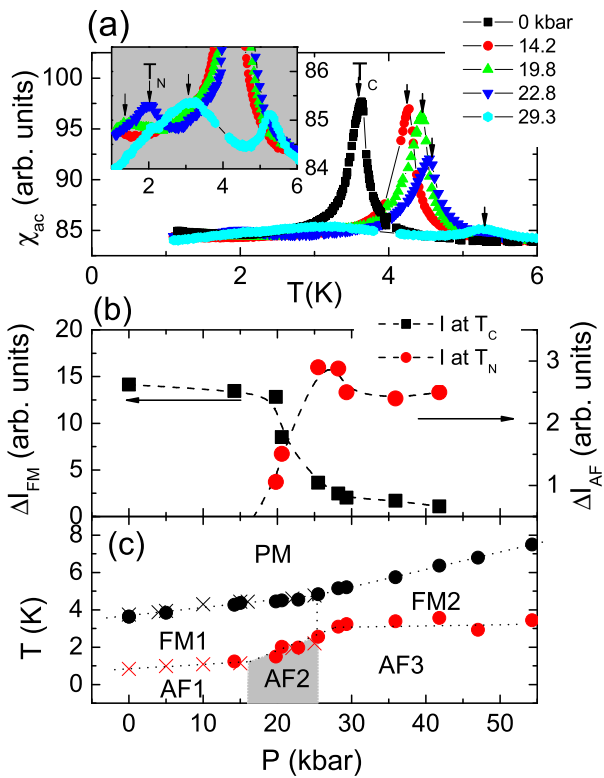}
\caption{(color online). (a) AC magnetic susceptibility $\chi _{\text{ac}}$ at 0, 14.2, 19.8, 22.8, and 29.3~kbar for $H_{ac} //$ab-plane. Inset: $\chi _{\text{ac}}$ versus $T$ near $T_{N}$ transition point. (b) Peak intensities both at $T_c$ and $T_N$, $\Delta I_{FM,AFM} = I (T_{c},T_N) - I (10K)$, are plotted as a function of pressure. (c) T-P phase diagram. FM and AFM transition temperatures are determined by both electrical resistivity (crosses) and magnetic susceptibility (circles). Dashed lines are guides to the eyes.}
\label{figure4}
\end{figure}
Figure~4(a) shows ac magnetic susceptibility as a function of temperature at several pressures for $H_{ac} \bot$c-axis. A sharp peak corresponding to the ferromagnetic transition occurs at 3.5~K for $P=0$~kbar and moves toward higher temperature with $P$ at the same rate as that of $T_{c}$ determined by the resistivity, as shown in Fig.~4(c). The resistivity anomaly at the N$\acute{e}$el temperature dramatically changes from Fisher-Langer (FL) or ferromagnetic like to Suezaki-Mori (SM) behavior with increasing pressure, i.e., from positive divergence to negative divergence in $d\rho /dT$ at $T_N$. Salamon claimed that it is prerequisite for magnetic superzone gaps that the ordering wavevector $K_A$ should connect portions of the Fermi surface. \cite{salamon72} In rare earth metals, spiral spin structures lead to small values of $K_A$ which span the Fermi surface, at least in some directions and, therefore, a SM behavior in resistivity. In beta brass where disorder makes it difficult for $K_A$ to span the Fermi surface, in contrast, FL behavior was reported.\cite{simons71} The inset to Fig.~4(a) magnifies the temperature range near $T_{N}$. At low pressures (AF1 in Fig.~4c), there is no clear signature in $\chi_{ac}$ corresponding to the AF transition below 1~K (not shown). Above 14.2~kbar (AF2 in Fig.~4c), a peak appears at the temperature corresponding to $T_{N}$ determined from $\rho$ and becomes enhanced with pressure (see inset of Fig.~4a and Fig.~4b), implying that the AF structure at intermediate pressures is not simple, but rather has a canted or helical structure. The concurrence of the SM behavior in the resistivity anomaly and the peak feature in $\chi_{\text{ac}}$ near 17~kbar is similar to the $\gamma$-phase Fe-Mn alloys, Fe$_x$Mn$_{1-x}$, where the resistivity anomaly change from FL type to SM type near $x=0.3$ was associated with a spin structure change from colinear to non-colinear one \cite{endoh71} and gap formation.\cite{asano71} Even though we need to determine exact spin structures from other measurements, such as neutron scattering under pressure, the above analogy suggests that it is only for intermediate pressures where the conditions for magnetic superzone gaps are met in CeZn$_{0.66}$Sb$_2$, thus leading to the SM-type resistivity anomaly. Magnetic field dependence of the anomaly is also consistent with the analysis in that the superzone gaps formed at $H=0$~Oe are destroyed with magnetic field at intermeidate pressures, thus leading to a change from the SM-type to the FL-type (see Fig.~2).

We have reported a pressure-induced, first-order resistivity anomaly where the resistivity of CeZn$_{0.66}$Sb$_{2}$ increases with decreasing temperature and shows thermal hysteresis at the antiferromangetic transition temperature $T_N$. By subtracting a low-pressure data set, we directly compared our experiments to Suezaki-Mori theory without any assumption on backgrounds and found reasonably good agreement both below and above $T_N$. The dramatic pressure effects on the resistivity anomaly, from the low-pressure Fisher-Langer type to the intermediate-pressure Suezaki-Mori type, are explained in terms of gap formation on the Fermi surface when the AF ordering wavevector $K_A$ is tuned to span the Fermi surface by pressure. Magnetic field dependence of the anomaly is also consistent with gap formation at intermediate pressures. 

Work at Los Alamos was performed under the auspices of the U.S. Department of Energy/Office of Science. Work at UC-Davis was supported by by NSF Grant No. DMR-0433560. V.A.S. acknowledges the support of Russian Foundation for Basic Research (Grant No. 03-02-17119) and Program ``Physics and Mechanics of Strongly Compressed Matter of Presidium of Russian Academy of Sciences''. We benefited from stimulating discussion with M. B. Salamon.

\bibliography{CeZnSb2}

\begin{thebibliography}{18}
\expandafter\ifx\csname natexlab\endcsname\relax\def\natexlab#1{#1}\fi
\expandafter\ifx\csname bibnamefont\endcsname\relax
  \def\bibnamefont#1{#1}\fi
\expandafter\ifx\csname bibfnamefont\endcsname\relax
  \def\bibfnamefont#1{#1}\fi
\expandafter\ifx\csname citenamefont\endcsname\relax
  \def\citenamefont#1{#1}\fi
\expandafter\ifx\csname url\endcsname\relax
  \def\url#1{\texttt{#1}}\fi
\expandafter\ifx\csname urlprefix\endcsname\relax\def\urlprefix{URL }\fi
\providecommand{\bibinfo}[2]{#2}
\providecommand{\eprint}[2][]{\url{#2}}

\bibitem[{\citenamefont{Hill}(1970)}]{hill70}
\bibinfo{author}{\bibfnamefont{H.~H.} \bibnamefont{Hill}},
  \emph{\bibinfo{title}{Plutonium 1970 and Other Actinides}}
  (\bibinfo{publisher}{AIME, New York}, \bibinfo{year}{1970}),
  p.~\bibinfo{pages}{2}.

\bibitem[{\citenamefont{Fisk et~al.}(1988)\citenamefont{Fisk, Hess, Pethick,
  Pines, Smith, Thompson, and Willis}}]{fisk88}
\bibinfo{author}{\bibfnamefont{Z.}~\bibnamefont{Fisk}},
  \bibinfo{author}{\bibfnamefont{D.~W.} \bibnamefont{Hess}},
  \bibinfo{author}{\bibfnamefont{C.~J.} \bibnamefont{Pethick}},
  \bibinfo{author}{\bibfnamefont{D.}~\bibnamefont{Pines}},
  \bibinfo{author}{\bibfnamefont{J.~L.} \bibnamefont{Smith}},
  \bibinfo{author}{\bibfnamefont{J.~D.} \bibnamefont{Thompson}},
  \bibnamefont{and} \bibinfo{author}{\bibfnamefont{J.~O.}
  \bibnamefont{Willis}}, \bibinfo{journal}{Science}
  \textbf{\bibinfo{volume}{239}}, \bibinfo{pages}{33} (\bibinfo{year}{1988}),
  \bibinfo{note}{references therein}.

\bibitem[{\citenamefont{Thompson}(1992)}]{thompson92}
\bibinfo{author}{\bibfnamefont{J.~D.} \bibnamefont{Thompson}},
  \emph{\bibinfo{title}{Selected Topics in Magnetism}}
  (\bibinfo{publisher}{World Scientific}, \bibinfo{year}{1992}),
  vol.~\bibinfo{volume}{2}, p. \bibinfo{pages}{107}.

\bibitem[{\citenamefont{Doniach}(1977)}]{doniach77}
\bibinfo{author}{\bibfnamefont{S.}~\bibnamefont{Doniach}},
  \emph{\bibinfo{title}{Valence Instabilities and Related Narrow-Band
  Phenomena}} (\bibinfo{publisher}{Plenum, New York}, \bibinfo{year}{1977}), p.
  \bibinfo{pages}{169}.

\bibitem[{\citenamefont{Mackintosh}(1962)}]{mackintosh62}
\bibinfo{author}{\bibfnamefont{A.~R.} \bibnamefont{Mackintosh}},
  \bibinfo{journal}{Phys. Rev. Lett.} \textbf{\bibinfo{volume}{9}},
  \bibinfo{pages}{90} (\bibinfo{year}{1962}).

\bibitem[{\citenamefont{Salamon}(1972)}]{salamon72}
\bibinfo{author}{\bibfnamefont{M.~B.} \bibnamefont{Salamon}},
  \bibinfo{journal}{J. Phys. C: Solid St. Phys.} \textbf{\bibinfo{volume}{5}},
  \bibinfo{pages}{L31} (\bibinfo{year}{1972}).

\bibitem[{\citenamefont{Suezaki and Mori}(1969)}]{suezaki69}
\bibinfo{author}{\bibfnamefont{Y.}~\bibnamefont{Suezaki}} \bibnamefont{and}
  \bibinfo{author}{\bibfnamefont{H.}~\bibnamefont{Mori}},
  \bibinfo{journal}{Prog. Theor. Phys.} \textbf{\bibinfo{volume}{41}},
  \bibinfo{pages}{1177} (\bibinfo{year}{1969}).

\bibitem[{\citenamefont{Craven and Parks}(1973)}]{craven73}
\bibinfo{author}{\bibfnamefont{R.~A.} \bibnamefont{Craven}} \bibnamefont{and}
  \bibinfo{author}{\bibfnamefont{R.~D.} \bibnamefont{Parks}},
  \bibinfo{journal}{Phys. Rev. Lett.} \textbf{\bibinfo{volume}{31}},
  \bibinfo{pages}{383} (\bibinfo{year}{1973}), \bibinfo{note}{references
  therein}.

\bibitem[{\citenamefont{Thomas et~al.}(1973)\citenamefont{Thomas, Giray, and
  Parks}}]{thomas73}
\bibinfo{author}{\bibfnamefont{G.~A.} \bibnamefont{Thomas}},
  \bibinfo{author}{\bibfnamefont{A.~B.} \bibnamefont{Giray}}, \bibnamefont{and}
  \bibinfo{author}{\bibfnamefont{R.~D.} \bibnamefont{Parks}},
  \bibinfo{journal}{Phys. Rev. Lett.} \textbf{\bibinfo{volume}{31}},
  \bibinfo{pages}{241} (\bibinfo{year}{1973}).

\bibitem[{\citenamefont{Lee}(2005)}]{hanoh05}
\bibinfo{author}{\bibfnamefont{H.}~\bibnamefont{Lee}} (\bibinfo{year}{2005}),
  \bibinfo{note}{unpublished}.

\bibitem[{\citenamefont{Asai et~al.}(2003)\citenamefont{Asai, Thamizhavel,
  Shishido, Ueda, Inada, Settai, Kobayashi, and Onuki}}]{asai03}
\bibinfo{author}{\bibfnamefont{R.}~\bibnamefont{Asai}},
  \bibinfo{author}{\bibfnamefont{A.}~\bibnamefont{Thamizhavel}},
  \bibinfo{author}{\bibfnamefont{H.}~\bibnamefont{Shishido}},
  \bibinfo{author}{\bibfnamefont{T.}~\bibnamefont{Ueda}},
  \bibinfo{author}{\bibfnamefont{Y.}~\bibnamefont{Inada}},
  \bibinfo{author}{\bibfnamefont{R.}~\bibnamefont{Settai}},
  \bibinfo{author}{\bibfnamefont{T.~C.} \bibnamefont{Kobayashi}},
  \bibnamefont{and} \bibinfo{author}{\bibfnamefont{Y.}~\bibnamefont{Onuki}},
  \bibinfo{journal}{J. Phys.: Condens. Matter} \textbf{\bibinfo{volume}{15}},
  \bibinfo{pages}{L463} (\bibinfo{year}{2003}).

\bibitem[{\citenamefont{Zumsteg and Parks}(1970)}]{zumsteg70}
\bibinfo{author}{\bibfnamefont{F.~C.} \bibnamefont{Zumsteg}} \bibnamefont{and}
  \bibinfo{author}{\bibfnamefont{R.~D.} \bibnamefont{Parks}},
  \bibinfo{journal}{Phys. Rev. Lett.} \textbf{\bibinfo{volume}{24}},
  \bibinfo{pages}{520} (\bibinfo{year}{1970}).

\bibitem[{\citenamefont{Kadanoff et~al.}(1967)\citenamefont{Kadanoff, Gotze,
  Hamblen, Hecht, Lewis, Palciauskas, Rayl, Swift, Aspnes, and
  Kane}}]{kadanoff67}
\bibinfo{author}{\bibfnamefont{L.~P.} \bibnamefont{Kadanoff}},
  \bibinfo{author}{\bibfnamefont{W.}~\bibnamefont{Gotze}},
  \bibinfo{author}{\bibfnamefont{D.}~\bibnamefont{Hamblen}},
  \bibinfo{author}{\bibfnamefont{R.}~\bibnamefont{Hecht}},
  \bibinfo{author}{\bibfnamefont{E.~A.~S.} \bibnamefont{Lewis}},
  \bibinfo{author}{\bibfnamefont{V.~V.} \bibnamefont{Palciauskas}},
  \bibinfo{author}{\bibfnamefont{M.}~\bibnamefont{Rayl}},
  \bibinfo{author}{\bibfnamefont{J.}~\bibnamefont{Swift}},
  \bibinfo{author}{\bibfnamefont{D.}~\bibnamefont{Aspnes}}, \bibnamefont{and}
  \bibinfo{author}{\bibfnamefont{J.}~\bibnamefont{Kane}},
  \bibinfo{journal}{Rev. Mod. Phys.} \textbf{\bibinfo{volume}{39}},
  \bibinfo{pages}{395} (\bibinfo{year}{1967}).

\bibitem[{\citenamefont{Fisher and Langer}(1968)}]{fisher68}
\bibinfo{author}{\bibfnamefont{M.~E.} \bibnamefont{Fisher}} \bibnamefont{and}
  \bibinfo{author}{\bibfnamefont{J.~S.} \bibnamefont{Langer}},
  \bibinfo{journal}{Phys. Rev. Lett.} \textbf{\bibinfo{volume}{20}},
  \bibinfo{pages}{665} (\bibinfo{year}{1968}).

\bibitem[{\citenamefont{Akiba and Mitsui}(1972)}]{akiba72}
\bibinfo{author}{\bibfnamefont{C.}~\bibnamefont{Akiba}} \bibnamefont{and}
  \bibinfo{author}{\bibfnamefont{T.}~\bibnamefont{Mitsui}},
  \bibinfo{journal}{J. Phys. Soc. Jpn.} \textbf{\bibinfo{volume}{32}},
  \bibinfo{pages}{644} (\bibinfo{year}{1972}).

\bibitem[{\citenamefont{Simons and Salamon}(1971)}]{simons71}
\bibinfo{author}{\bibfnamefont{D.~S.} \bibnamefont{Simons}} \bibnamefont{and}
  \bibinfo{author}{\bibfnamefont{M.~B.} \bibnamefont{Salamon}},
  \bibinfo{journal}{Phys. Rev. Lett.} \textbf{\bibinfo{volume}{26}},
  \bibinfo{pages}{750} (\bibinfo{year}{1971}).

\bibitem[{\citenamefont{Endoh and Ishikawa}(1971)}]{endoh71}
\bibinfo{author}{\bibfnamefont{Y.}~\bibnamefont{Endoh}} \bibnamefont{and}
  \bibinfo{author}{\bibfnamefont{Y.}~\bibnamefont{Ishikawa}},
  \bibinfo{journal}{J. Phys. Soc. Jpn.} \textbf{\bibinfo{volume}{30}},
  \bibinfo{pages}{1614} (\bibinfo{year}{1971}).

\bibitem[{\citenamefont{Asano and Yamashita}(1971)}]{asano71}
\bibinfo{author}{\bibfnamefont{S.}~\bibnamefont{Asano}} \bibnamefont{and}
  \bibinfo{author}{\bibfnamefont{J.}~\bibnamefont{Yamashita}},
  \bibinfo{journal}{J. Phys. Soc. Jpn.} \textbf{\bibinfo{volume}{31}},
  \bibinfo{pages}{1000} (\bibinfo{year}{1971}).

\end{thebibliography}

\end{document}